\documentstyle[11pt]{article}
\newcommand{\be}{\begin{equation}}
\newcommand{\ee}{\end{equation}}
\newcommand{\bea}{\begin{eqnarray}}
\newcommand{\eea}{\end{eqnarray}}

\newcommand{\vs}[1]{\vspace{#1 mm}}
\newcommand{\hs}[1]{\hspace{#1 mm}}

\begin{document}
\thispagestyle{empty}

\rightline{hep-th/0206057}

\vs{20}

\centerline{\large\bf Intersecting S-Brane Solutions of $D=11$ Supergravity} 
\vs{15}
\centerline{N.S. Deger\footnote{e-mail: deger@gursey.gov.tr} and A. 
Kaya\footnote{e-mail: kaya@gursey.gov.tr}}
\vs{5}
\centerline{Feza Gursey Institute, Emek Mah. No:68, 81220, Cengelkoy,
Istanbul, TURKEY}
\vs{25}

\begin{abstract}
We construct all possible orthogonally intersecting S-brane solutions in
11-dimensions corresponding to standard supersymmetric M-brane intersections.
It is found that the solutions can be obtained by 
multiplying the brane and the transverse directions
with appropriate powers of two hyperbolic functions of
time. This is the S-brane analog of the ``harmonic function
rule''. The transverse directions can be hyperbolic, flat or
spherical. We also discuss some properties of these solutions. 
\end{abstract} 

\newpage

\setcounter{page}{1}

\section{Introduction}

\hs{5}D-branes \cite{polc} have played a very important role 
in our understanding of
non-perturbative aspects of string/M theory and recently the AdS/CFT
duality \cite{ads1, ads2, ads3}. As is well known, in perturbative string
theory at weak string coupling, D-branes can be described as hypersurfaces
where open strings can end. This is achieved by imposing Dirichlet
boundary conditions along transverse spacelike directions in the string
world-sheet action. On the other hand, D-branes have an alternative
description as soliton solutions of low energy supergravity
theories. These solutions are important in determining the holographic
properties of D-branes. Having two different pictures for D-branes is the
essence of the AdS/CFT duality.

\
\

Naturally, in string perturbation theory, one can also consider open
strings obeying Dirichlet boundary conditions along time-like or null
directions. These are space-like or null analogs of D-branes and are
called S-branes \cite{s1} or N-branes \cite{kogan},  respectively. $Sp$-branes 
can be viewed as
spacelike $p$-branes (with $p+1$ spatial dimensions) which exist only
for a moment in time. In world-sheet conformal field theory, their
description is very similar to usual D-branes. For instance, it is
possible to describe S-branes by a boundary state. S-branes can
also be considered as time dependent tachyonic kink solutions of the
unstable D-brane world-volume theories \cite{sen}.

\
\

S-branes are expected to play the role of D-branes in realizing $dS/CFT$
duality \cite{strominger} in string/M theory. As D-branes lead us to
holography in space-like directions, one hopes that S-branes will imply
time-like holography. Moreover, formulating string theory on
time dependent backgrounds is important in searching cosmological
applications of the theory  (for some recent developments see for example
\cite{time1}-\cite{time5}). 

\
\

If S-branes are stable objects (even though they 
are not supersymmetric), then one should be able to construct corresponding
supergravity solutions. In \cite{hull1, hull2} timelike T-dual 
of type II string theories were
considered and Euclidean brane solutions were constructed. This can be
generalized to usual type II and M theories as was done in \cite{s1,
s2, s3} (see also \cite{yeni0}-\cite{s4} for some earlier work). 
In \cite{s1} S0-brane solution of $D=4$ 
Einstein-Maxwell theory and SM5-brane
solution of $D=11$ supergravity were found. Later, all SD-brane solutions
were constructed in \cite{s2} and \cite{s3}. In \cite{s2} these were
obtained by solving Einstein's equations for a dilaton, gravity and an
arbitrary rank antisymmetric tensor field system in an arbitrary
dimension. Whereas in \cite{s3}
solution generating techniques are applied to an appropriate
time-dependent solution of the 11-dimensional vacuum Einstein equations.
The connection between solutions of \cite{s2} and \cite{s3} is shown in
\cite{s5}.

\
\

Following these developments, a natural direction of study is to
search for supergravity solutions of intersecting spacelike brane
configurations. Intersecting solutions enriched our understanding
of $AdS/CFT$ duality and something similar is likely to happen in the
context of $dS/CFT$ correspondence too. 
In \cite{s3}, multiply charged SD$p$/SD($p-2$) and
SD$p$/SD($p-4$) brane solutions are constructed.
In this paper, we will construct all possible orthogonally
intersecting S-brane solutions of $D=11$ supergravity theory
corresponding to supersymmetric M-brane intersections. Our main
observation is that all solutions can be obtained by a simple
procedure of multiplying the brane and the transverse directions by
two hyperbolic functions of time. This is the S-brane analog of the 
harmonic function rule \cite{har1, har2}. 

\
\

The organization of the paper is as follows; in the
next section we give the main construction rules for intersecting
S-brane solutions. In section 3, using these rules we write down the solutions
explicitely and discuss their properties.  We conclude in section
4. Technical details are presented in the appendix.

\section{Basic rules}

\hs{5}In this section 
we will consider all possible orthogonal intersections of SM2 and
SM5-branes in 11-dimensions corresponding to standard supersymmetric
M-brane intersections \cite{har1, har2, har3}. As we will comment, other 
cases turn out to
have technical difficulties and cannot be obtained by the rules of this
section. Similar to the usual brane
intersections, one can define {\it common tangent}, {\it relative
transverse} and {\it overall transverse} directions in an obvious
way. Following the convention in the literature, an S$p$ brane is
defined to have $p+1$-dimensional spatial world-volume. For each 
brane, we introduce three constants $q_i$, $t_i$ and $M_i$,
where $q_i$ is the charge, $t_i$ is an instant in time and $M_i$ is a
positive  number.  We will comment on the 
physical interpretations of
$t_i$ and $M_i$ in the next section. Let $n+1$ be the dimension of the overall
transverse directions including time. We take the $n$-dimensional
transverse spatial hypersurface to be $\Sigma_{n,\sigma}$, that is 
the unit sphere, the unit hyperbola or flat space when $\sigma=1$,
$\sigma=-1$ and $\sigma=0$, respectively. The 11-dimensional space can
be parametrized by the coordinates
$(x_1,..,x_p,y_1,..y_q,t,\Sigma_{n,\sigma})$, where $x$ and $y$ 
parametrize the common tangent and relative transverse
directions and $(t,\Sigma_n)$ is the overall transverse space. We
consider a diagonal metric where the metric functions depend only on $t$.  

\
\

For each brane we define the following function which depends on
the three constants we introduced
\be\label{H}
H_i=\frac{q_i^2}{M_i^2}\,\cosh^2\,\left[\,M_i\,(t-t_i)\,\right].
\ee
In the construction, we will need one more function which
characterizes the transverse space
\be\label{G}
G_{n,\sigma}=\begin{cases}
{\frac{2n(n-1)}{M^2}\sinh^2\left[M\sqrt{\frac{(n-1)}{2n}}\,\,t\right]\hs{5}
\sigma=-1 \,\,\,\,\textrm{(hyperbola)},\cr\cr  
\frac{2n(n-1)}{M^2}\cosh^2\left[M\sqrt{\frac{(n-1)}{2n}}\,\,t\right] \hs{5}
\sigma=1\,\,\,\,\textrm{(sphere)},\cr\cr
e^{2M\sqrt{\frac{(n-1)}{2n}}\,\,t} \hs{31}\sigma=0 \,\,\,\,\textrm{(flat)},}
\end{cases}
\ee
where 
\be\label{sum}
M^2=\sum_i M_i^2.
\ee
In terms of these functions, the metric corresponding to any number of
possible standard  intersection of SM-branes can be obtained by the
following simple rules. Up to the $H$-functions, the
overall transverse space takes the form
\be\label{overall}
G_{n,\sigma}^{-\frac{n}{(n-1)}}\left[\,-dt^2\,+\,G_{n,\sigma}\,d\Sigma_{n,\sigma}^2\,\right] 
\ee
where $d\Sigma_{n,\sigma}^2$ is the metric on the unit sphere,
the unit hyperbola or flat space. One can find the dependence of the
metric on $H$-functions simply by multiplying the brane and the
transverse directions for $i$'th brane by appropriate powers of
$H_i$ which are 
\be\label{m2h}
\textrm{For i'th SM2-brane:}\begin{cases}
{\textrm{Brane direction}\hs{12}H_i^{-1/3}\cr\cr
 \textrm{Transverse direction}\hs{5}H_i^{1/6}}\end{cases}
\ee
\be\label{m5h}
\textrm{For j'th SM5-brane:}\begin{cases}
{\textrm{Brane direction}\hs{12}H_j^{-1/6}\cr\cr
 \textrm{Transverse direction}\hs{5}H_j^{1/3}}\end{cases}
\ee
In other words, each $H$ function appears in the metric as
$H^{1/2}$ multiplying the directions transverse to that brane and
there is an overall conformal factor with $H^{-1/3}$ for SM2-brane
and $H^{-1/6}$ for SM5-brane. This is very similar to the harmonic
function rule of supersymmetric M-branes \cite{har1,har2}. Note that
all these are valid when $n\geq2$, and $n=0,1$ cases are degenerate.    

\
\

After determining the metric, let us now describe the form of the
anti-symmetric tensor fields. The transverse space (including time) to
any SM2-brane is 8-dimensional. Let $\textrm{Vol}_7$ be the closed
volume-form of the 7-dimensional spatial transverse space, where it is
defined by ignoring the $H$ and the $G$ functions in the
metric. Generically, the volume-form can be written as
$\textrm{Vol}_7=dy_1\wedge dy_2 \wedge..\wedge dy_{7-n}\wedge
\Omega_n$, where $\Omega_n$ is the volume form on
$\Sigma_{n,\sigma}$.  Then the four-form field of the $i$'th SM2-brane can be
written as  
\be\label{fm2}
F=q_i\,\,*\,\textrm{Vol}_7,
\ee
where $*$ is the Hodge dual with respect to the full metric. Similarly,
for each SM5-brane, the space-like transverse space is
4-dimensional. Defining $\textrm{Vol}_4$ to be the volume-form of this space
without the $H$ and $G$ functions, then the four form field
corresponding to the j'th SM5-brane is equal to
\be\label{fm5}
F=q_j\,\,\textrm{Vol}_4.
\ee
It is easy to see that we have $dF=0$ and $d*F=0$ for both cases. 

\
\

Which type of intersections are allowed for
SM-branes? Since $n=0,1$ cases are {\it degenerate},
the overall transverse space
(including time) should at least be 3-dimensional. Using $Sp\perp
Sq(r)$ to denote intersection of  S$p$ and S$q$ branes over an S$r$ brane,
the possible double intersections are 
\bea
SM2\perp SM2(0)\nonumber\\
SM2\perp SM5(1)\label{i2}\\
SM5\perp SM5(3)\nonumber
\eea
Note that the S-brane intersection corresponding to standard $M5 \perp M5(1)$  
has $n=0$ and thus it is degenerate.  For triple intersections, the
only possible cases are 
\bea
SM2\perp SM2 \perp SM2(0) \nonumber\\
SM2\perp SM2 \perp SM5(0) \label{i3}\\
SM2\perp SM5 \perp SM5(1) \nonumber 
\eea
The intersection of three SM5-branes over an SM3 brane has $n=0$ and
thus this configuration is degenerate. Finally, the only possible
quartic intersection is 
\bea
SM2\perp SM2 \perp SM5 \perp SM5(0) \label{i4}
\eea
This corresponds to the special intersection of 4
M-branes, where the solution preserves 1/8 supersymmetry and not 1/16
as one would naively expect.

\
\

Since S-branes do not preserve any supersymmetry, one may try to
consider other possible intersections. Let us first determine which
cases are consistent with our choice of the metric and the four-form
field (\ref{fm2}) and (\ref{fm5}). In 11-dimensions the
four-form field equations are $dF=0$ and $d*F\sim F\wedge F$. However, in
our ansatz we have $d*F=0$ and $dF=0$. Therefore any configuration
which has $F\wedge F\not=0$ cannot be written by using S-brane
rules. Another constraint comes from the fact that the Ricci tensor of
the metric is diagonal. In Einstein equations, there are terms 
coming from the $F_{AMNP}F_{B}{}^{MNP}$  contraction. Any brane intersection
giving non-diagonal contribution through this contraction cannot be
written by using S-brane rules. 

\
\

For instance, consider $SM5\perp SM5(1)$. Although this intersection is
degenerate with respect to S-brane rules, one may still insist on
finding a time dependent solution.  The space-time  is 
decomposed as (11=2+4+$4'$+1) and the four-form field is equal
$F=q_5\textrm{Vol}_{4'}+q_{5'}\textrm{Vol}_4$  and thus $F\wedge
F\not=0$. Therefore, to be able to satisfy the four-form field equations
one should consider a more general ansatz. Similarly, let us
also consider the intersection of two
SM5-branes on an S4-brane. Then the 11-dimensional space can be
decomposed into $(t,x_1,..,x_5,y_1,y_2,\Sigma_3)$, where
$(x_1,..,x_5,y_1)$ and $(x_1,..,x_5,y_2)$ are the coordinates for the
two SM5-branes, respectively. Then, the four-form field
becomes $F=q_5\, dy_1\wedge\Omega_3\,\,+\,\,q_{5'}\,dy_2\wedge\Omega_3$.
Although the four-form field equations are satisfied, it
is easy to see that in Einstein equations there appears a non-diagonal
contribution along $y_1y_2$ directions coming from the
$F_{AMNP}F_{B}{}^{MNP}$  contraction. This term cannot be canceled by the
Ricci tensor of the metric, which is diagonal in our ansatz.

\
\

Taking into account of these constraints, one still finds, for
instance, three more double intersections which are consistent with
our ansatz. These are $SM2\perp SM2(-1)$, $SM2\perp SM5(0)$, $SM5\perp
SM5(2)$. For these cases, we could not succeed in diagonalizing 
the differential equations as we did for the standard SM-brane
intersections. Therefore, it seems difficult to find explicit solutions.
This shows that, our S-brane rules can only be used to obtain standard
S-brane intersections. Let us also recall that, even for standard
cases, the overall transverse space (including time) should at least
be 3-dimensional.

\section{Solutions in 11-dimensions} 

\hs{5}In this section, we apply the rules given above to construct
intersecting solutions. 
Although single SM2 and SM5-brane solutions have been obtained
previously, we start with the construction of these  backgrounds for
completeness. Let us first consider the single SM2-brane solution. In
the presence of an SM2-brane, the 
11-dimensional space-time can be decomposed as (11=3+1+7). As discussed in
the previous section, we introduce 3-constants $q$, $t_0$ and
$M$. We also have $n=7$. Then using (\ref{overall}), (\ref{m2h}) and
(\ref{fm2}), the solution can be written as
\bea\label{sl1}
ds^2&=&H^{-1/3}\,\left[dx_1^2+dx_2^2+dx_3^2\right]\,+\,H^{1/6}\,G_{7,\sigma}^{-7/6}\,\left[-dt^2\,+\,G_{7,\sigma}\,d\Sigma_{7,\sigma}^2\right]\\
F&=&q*\Omega_7,
\eea
where $H$ and $G$ are given in (\ref{H}) and (\ref{G}),
respectively. Note that the Hodge dual $*$ is defined with respect to
the full metric (\ref{sl1}).

\
\

Similarly, one can consider a single SM5-brane configuration. This
time we have (11=6+1+4)  splitting of space-time and $n=4$. From 
(\ref{overall}), (\ref{m5h}) and (\ref{fm5}), we obtain
\bea
ds^2&=&H^{-1/6}\,\left[dx_1^2+..+dx_6^2\right]\,+\,H^{1/3}\,G_{4,\sigma}^{-4/3}\,\left[-dt^2\,+\,G_{4,\sigma}\,d\Sigma_{4,\sigma}^2\right]\\
F&=&q\,\Omega_4
\eea
where $H$ and $G$ are again given in (\ref{H}) and (\ref{G}),
respectively. 

\
\

We now consider $SM2\perp SM2(0)$ intersection.  We have $n=5$ and the
space-time is decomposed as (11=1+2+2+1+5). The solution is specified by
6-constants $(q_1,t_1,M_1)$ and $(q_2,t_2,M_2)$ corresponding to the
first and the second SM2-branes, respectively. Using S-brane rules of
the previous section we obtain
\bea
ds^2&=&(H_1H_2)^{-1/3}\left[dx^2\right]\,+\,H_1^{-1/3}H_2^{1/6}\left[dy_1^2+dy_2^2\right]\,+\,H_{1}^{1/6}H_{2}^{-1/3}\left[dy_3^2+dy_4^2\right]\nonumber\\
&+&(H_1H_2)^{1/6}\,G_{5,\sigma}^{-5/4}\left[-dt^2\,+\,G_{5,\sigma}\,d\Sigma_{5,\sigma}^2\right],\label{16}\\
F&=&q_1*(dy_3\,dy_4\,\Omega_5)\,\,+\,\,q_2*(dy_1\,dy_2\,\Omega_5).
\eea
The first SM2-brane has the charge $q_1$ and world-volume coordinates
$(x,y_1,y_2)$. This brane is characterized by the
function $H_1(q_1,t_1,M_1)$ given in (\ref{H}). Similarly the second
SM2-brane has the charge $q_2$, world-volume $(x,y_3,y_4)$ and the
function $H_2(q_2,t_2,M_2)$. 

\
\

Construction of $SM5\perp SM5(3)$ runs along similar lines. The
space-time is decomposed as (11=4+2+2+1+2) and $n=2$. There are again
6-constants, 3 for each SM5-brane. Using S-brane rules one finds
\bea
ds^2&=&(H_1H_2)^{-1/6}\left[dx_1^2+..+dx_4^2\right]\,+\,
H_1^{-1/6}H_2^{1/3}\left[dy_1^2+dy_2^2\right]\,+\,H_{1}^{1/3}H_{2}^{-1/6}\left[dy_3^2+dy_4^2\right]\nonumber\\
&+&(H_1H_2)^{1/3}\,G_{2,\sigma}^{-2}\left[-dt^2\,+\,G_{2,\sigma}\,d\Sigma_{2,\sigma}^2\right],\\
F&=&q_1 (dy_3\,dy_4\,\Omega_2)\,\,+\,\,q_2 (dy_1\,dy_2\,\Omega_2).
\eea
The first SM5-brane has the charge $q_1$ and oriented along
$(x_1,..,x_4,y_1,y_2)$ hyperplane. Similarly, the second SM5-brane has
the charge $q_2$ and world-volume directions $(x_1,..,x_4,y_3,y_4)$. 

\
\

Let us now obtain $SM2\perp SM5(1)$ which is the last possible double
intersection. The space-time is decomposed as (11=2+1+4+1+3). Let $q_1$
and $q_2$ be the charges of the SM2 and SM5-branes respectively. Then
the solution can be written as 
\bea
ds^2&=&H_1^{-1/3}H_2^{-1/6}\left[dx_1^2+dx_2^2\right]\,+\,
H_1^{-1/3}H_2^{1/3}\left[dy_1^2\right]\,+\,H_{1}^{1/6}H_{2}^{-1/6}
\left[dy_2^2+..+dy_5^2\right]\nonumber\\
&+&H_1^{1/6}H_2^{1/3}\,G_{3,\sigma}^{-3/2}\left[-dt^2\,+\,G_{3,\sigma}\,d\Sigma_{3,\sigma}^2\right],\\
F&=&q_1 *(dy_2..dy_5\,\Omega_3)\,\,+\,\,q_2 (dy_1\,\Omega_3).
\eea
The SM2-brane is oriented along $(x_1,x_2,y_1)$, and SM5-brane is
oriented along $(x_1,x_2,y_2,..,y_5)$ hyperplanes. Again the solution
depends on 6-constants, 3 for each SM-brane. 

\
\

The other intersecting solutions can easily be obtained once the brane
directions are specified. For $SM2\perp SM2\perp SM2(0)$ intersection let
$(x,y_1,y_2)$, $(x,y_3,y_4)$ and $(x,y_5,y_6)$ parametrize the three SM2-brane
world-volumes. So, $x$ is the common tangent direction and the overall
transverse space becomes $(t,\Sigma_3)$. The corresponding metric and
four-form field are
\bea
ds^2&=&(H_1H_2H_3)^{-1/3}\left[dx^2\right]\,+\,H_1^{-1/3}(H_2H_3)^{1/6}\left[dy_1^2+dy_2^2\right]\,+\,H_{2}^{-1/3}(H_1H_3)^{1/6}\left[dy_3^2+dy_4^2\right]\nonumber\\
&+& H_{3}^{-1/3}(H_1H_2)^{1/6}\left[dy_5^2+dy_6^2\right]+ (H_1H_2H_3)^{1/6}\,G_{3,\sigma}^{-3/2}\left[-dt^2\,+\,G_{3,\sigma}\,d\Sigma_{3,\sigma}^2\right],\\
F&=&q_1*(dy_3\,dy_4\,dy_5\,dy_6\,\Omega_3)\,\,+\,\,q_2*(dy_1\,dy_2\,dy_5\,dy_6\,\Omega_3)+q_3*(dy_1\,dy_2\,dy_3\,dy_4\,\Omega_3).
\eea
Note that the solution depends on 9 arbitrary constants, 3 for each SM2-brane.

\
\

Finally, let us construct the quartic intersection $SM2\perp SM2\perp SM5\perp
SM5(0)$. As we will see, the remaining two cases in (\ref{i3}) can be
obtained from this solution. We place two SM5-branes along
$(x,y_1,y_2,y_3,y_4,y_5)$ and $(x,y_1,y_2,y_3,y_6,y_7)$ hyperplanes. Over this
intersection, we add two SM2-branes oriented along $(x,y_4,y_6)$ and  
$(x,y_5,y_7)$ directions. The corresponding solution can be written as
\bea
ds^2&=&(H_1H_2)^{-1/3}(H_3H_4)^{-1/6}\left[dx^2\right]+(H_1H_2)^{1/6}(H_3H_4)^{-1/6}\left[dy_1^2+dy_2^2+dy_3^2\right]\nonumber\\
& + & H_1^{-1/3}H_2^{1/6}H_3^{-1/6}H_4^{1/3}\left[dy_4^2\right]+ 
      H_1^{1/6}H_2^{-1/3}H_3^{-1/6}H_4^{1/3}\left[dy_5^2\right]\nonumber\\
&+&   H_1^{-1/3}H_2^{1/6}H_3^{1/3}H_4^{-1/6}\left[dy_6^2\right]+ 
      H_1^{1/6}H_2^{-1/3}H_3^{1/3}H_4^{-1/6}\left[dy_7^2\right]\nonumber\\
& + & (H_1H_2)^{1/6}(H_3H_4)^{1/3}G_{2,\sigma}^{-2}\left[-dt^2\,+\,G_{2,\sigma}\,d\Sigma_{2,\sigma}^2\right],\\
F & = &
q_1*(dy_1\,dy_2\,dy_3\,dy_5\,dy_7\,\Omega_2)\,\,+\,\,q_2*(dy_1\,dy_2\,dy_3\,dy_4\,dy_6\,\Omega_2)\nonumber\\
&+&q_3 \,(dy_6dy_7\Omega_2)\,+\,q_4\,(dy_4dy_5\Omega_2).
\eea
This solution has 12 free parameters. 

\
\

In all these intersections, it is possible to
remove any number of brane configurations from the system. For
$i$'th brane if one considers the limit
\be\label{lim}
q_i\to 0,\hs{3}M_i\to 0,\hs{5}\frac{q_i}{M_i}=1,
\ee
then the $H$-function corresponding to this brane becomes
$H_i=1$. Thus all information about this brane dissappears,
giving a background with one less number of
intersections. In this way, one can obtain  $SM2\perp SM2\perp SM5(0)$
and $SM2\perp SM5\perp SM5(1)$ intersections from the quartic
intersection above.\footnote{However, note that not all
intersections can be obtained by this procedure since it may result
some smearing along overall transverse directions. For instance,
removing a brane from $SM2\perp SM2(0)$ gives the smeared version of
(\ref{sl1}).} Continuing like this, the flat space limit can
only be obtained when $\sigma=0,-1$ by applying the limit
(\ref{lim}) to each brane. In the  $\sigma=-1$ case, the
limit (\ref{lim}) gives the flat metric written in Rindler coordinates i.e
$-dt^2+t^2d\Sigma_{n,-1}$. For $\sigma=1$, the flat space limit is
singular.

\
\

There are 3-constants of integration $(q_i,t_i,M_i)$ for 
each brane. It turns out that it is possible to eliminate
one of the constants $M_i$ (for instance, by setting it's value to 1).
In all cases, this can be achieved by a 
scaling $t\to t/M_i$ followed by redefinitions $M_j\to M_iM_j$,
$j\not=i$.  Then by (\ref{sum}), $M$ should also be scaled as
$M\to M_iM$ so that $M^2=1+\sum_{j\not=i}M_j^2$. These should be
supplemented by further scalings of $x$ and/or $y$ coordinates or by
redefinitions of charges $q_i$, if necessary. 
For example, for single SM2-brane
and SM5-brane solutions  $x\to x M_i^{-1/3}$ and $x\to x M_i^{-1/6}$
scalings should also be performed, respectively. Similarly, in
$SM2\perp SM5(1)$ and $SM2\perp SM2\perp SM2(0)$ intersections, $x\to
M^{-1/2}_i$ and $x\to x/M_i$ scalings are enough to eliminate $M_i$ in
the solutions, respectively . In $SM2\perp SM2(0)$ intersection, in
addition to $x\to M^{-1/2}_ix$ scaling, $y\to M^{-1/8}_iy$ and $q\to
M_i^{1/4}q$ replacements  should also be performed. In all other cases,
one can find  similar scalings which leave the solutions invariant
and eliminate one of the constants $M_i$. Note however that, after
eliminating $M_i$ it is not anymore possible to remove that brane from 
the system. In particular the flat space limit cannot be taken when
$\sigma=0,-1$.  

\
\

What is the physical interpretation of these constants? It is clear that
$q_i$ is the electric or magnetic charge of the $i$'th brane. Since
S-branes are spacelike $p$-branes which exist only for an instant in
time, presumably $t_i$ can be identified with (or somehow be related to) 
that instant.\footnote{When $t_i$'s are different than each other,
then it might be more appropriate to call these solutions as
{\it overlapping} S-branes.} 
Note however that the solutions are not invariant under
$t\to t+ constant$. This suggests that S-branes are naturally
defined on a vacuum where there is no time translation
symmetry. 
Finally, it is tempting to claim that $M_i$ is related to
the energy of the $i$'th SM-brane. Note that since S-branes are not
supersymmetric, their masses can in principle be independent of their
charges (This is similar to having any number of electrons and protons
with charge +1 with arbitrary mass). Therefore, in S-brane
solutions, there should also be a free parameter describing the energy of
the brane. One support for identification of $M_i$ with the energy
comes from the observation that by a scaling $t\to t/M_i$ (followed by
others as explained in the above paragraph), it is possible to set
$M_i=1$. 
The energy operator $E=i\partial_t$  scales as $E\to M_i
E$. Thus, if the $i$'th S-brane has energy $M_i$ with respect to the
original time $t$, then after scaling it should have energy 1, which
is the case. Although this argument suggests the identification, to
verify that $M_i$ corresponds to the energy of the $i$'th S-brane
requires a precise definition of mass as a conserved quantity for
these time dependent backgrounds.

\
\

As in the case of usual M-brane solutions it is possible to apply
dimensional reduction to these solutions to obtain intersecting
S-branes of type IIA theory \cite{har4, har2}. To be able to reduce 
$\sigma=\pm1$
solutions  along overall transverse directions one should first 
find a way of smearing spherical or hyperbolic directions. For $n=2$,
smearing out gives two flat directions which reduces to
the $\sigma=0$ case. Therefore, this procedure is meaningful only
when $n\geq3$. It is straightforward to verify that upto $H$ factors,
the overall transverse space (\ref{overall}) can be flattened by one
direction\footnote{It is possible to have 
$g_{zz}(t) \sim e^{ct}$, where $c$ is another constant. This adds 
one more free constant to the list. The constant $M$ defined in
(\ref{sum}) should also be modified appropriately.} using 
\be\label{sm}
G_{n,\pm}^{-\frac{n}{(n-1)}}\left[\,-dt^2\,+\,G_{n,\pm}\,d\Sigma_{n,\pm}^2\,\right]\to
dz^2+ G_{n-1,\pm}^{-\frac{(n-1)}{(n-2)}}\left[\,-dt^2\,+\,G_{n-1,\pm}\,d\Sigma_{n-1,\pm}^2\,\right].
\ee
This procedure can be repeated until $\Sigma_{\pm}$ becomes
2-dimensional. It is now possible to reduce the solution along $z$
coordinate to find a solution of type IIA theory.  Note that, without
smearing out, one can still perform a double dimensional reduction
along common tangent or relative transverse directions.

\
\

A double dimensional reduction of SM2 and  SM5-branes 
gives spacelike NS-string (SNS1) and
SD4-brane of type IIA theory, respectively. A reduction along
transverse directions would give SD2-brane and SNS5-brane solutions,
respectively. By applying dimensional reduction along an overall
transverse direction, we obtain an $SD2\perp SD2(0)$ solution from
$SM2\perp SM2(0)$ intersection. It is now possible to apply several T-duality
transformations to obtain a list of intersecting SD-branes of type IIA
and type IIB-theory. On the other hand, one can also reduce $SM2\perp
SM2(0)$ solution along the common or one of the relative transverse
directions and this gives $SNS1\perp SNS1(0)$ or $SNS1\perp SD2 (0)$
intersections, respectively. In this way one can in principle obtain
all possible spacelike backgrounds corresponding to standard brane
intersections.    

\
\

Let us now try to determine the singularities and the asymptotics
of the solutions. For finite $t$, all metric functions are regular and
non-zero except for $G_{n,-1}$ in (\ref{G}) 
which vanishes at $t=0$. However, as $t\to 0$, $G_{n,-1}\to t$ and it is easy
to see that by defining a new coordinate  $u\sim t^{-1/(n-1)}$ the 
metric asymptotes to the flat space. More precisely, along 
the overall transverse  directions, the metric becomes
$-du^2+u^2d\Sigma_{n,-1}$ and $u\to\infty$. Therefore, for $\sigma=-1$, 
$t\in(0,\infty)$ and near the first asymptotic region as
$t\to 0$, the metric becomes locally flat space in Rindler coordinates. For
$\sigma=0,1$ solutions , $t\in (-\infty,+\infty)$.  

\
\

For $\sigma=0$, as $t\to -\infty$, the metric along overall transverse space
becomes\footnote{In the following, we will ignore numerical
coefficients which do not affect the results.}
\be
ds^2_{n+1}\to-e^{-\alpha t}dt^2 + e^{-\beta t}d\Sigma_{n,0}^{2},
\ee
where $\alpha > \beta >0$ for all values of $M_i$. Defining
$u=e^{-(\alpha/2)\,t}$, one finds that
\be
ds^2_{n+1}\to-du^{2}+u^{2\beta/\alpha}d\Sigma_{n,0}^2,\hs{5} u\to\+\infty.
\ee
Therefore, $\Sigma_{n,0}$ expands and the curvatures become
smaller as $t\to-\infty$. This shows that there is no curvature
singularity associated with overall transverse directions. 

\
\

For $\sigma=1$, $t\to-\infty$ limit is the same with the
$t\to+\infty$ limit; and for all three cases ($\sigma=0,\pm1$),
$t\to+\infty$ asymptote is the same. In this limit, the metric along overall
transverse space takes the form
\be\label{asy}
ds^2_{n+1}\to\,-e^{-\gamma t}\,dt^2 \,+\, e^{\delta t}\,\,d\Sigma_{n,\sigma}^{2},
\ee
where $\gamma>|\delta|\geq 0$, and $\delta$ can be positive,
negative or zero depending on the constants $M_i$. For
single SM2 and SM5-branes, and for $SM2\perp SM2(0)$ intersection
$\delta>0$. For $SM2\perp SM2\perp SM2(0)$ and $SM2\perp SM5\perp SM5(1)$
intersections $\delta\leq0$. For $SM5\perp SM5(3)$ and $SM2\perp SM2\perp 
SM5(0)$
intersections $\delta<0$, and for $SM2\perp SM5(1)$ and $SM2\perp
SM2\perp SM5\perp SM5(0)$ intersections $\delta$ can be
positive, negative or zero depending on the constants $M_i$. Defining
$u=e^{-(\gamma/2)\,t}$, the metric becomes
\be
ds^2_{n+1}\to\,-du^{2}\,+\,u^{-2\delta/\gamma}\,\,d\Sigma_{n,\sigma}^2,\hs{5} u\to 0.
\ee
Thus, for $\delta<0$, $\Sigma_{n,\sigma}$ collapses and the geometry
becomes singular along these directions. Note that for some
intersections, the sign of $\delta$ depends on the choice of the
constants $M_i$, and depending on this sign, the asymptotic behavior
alters. Also, if one smears out one of the spherical or hyperbolic
directions using (\ref{sm}), then the coefficients of the exponents in
(\ref{asy}) will change. For instance, $\delta$ may switch sign for
fixed $M_i$ after smearing.

\
\

In all these solutions there are generic singularities
associated with common tangent or relative transverse
directions. Specifically, the common tangent directions are always multiplied
by negative powers of $H$. So, as $t\to\pm\infty$, the coefficient of
the flat metric along these directions vanishes and the 
geometry becomes singular.
Similarly, along relative transverse directions,  there are positive
and negative  powers of $H$ functions multiplying the flat
metric. Therefore, depending on the constants $M_i$, these directions
may collapse, expand  or stay flat as $t\to\pm\infty$. If they do
collapse, this implies existence of generic singularities associated with
relative transverse directions. 

\
\

As a final technical remark, let us note that if one defines the
asymptotic region where the radius of the hyperbola
or sphere diverges, then for $\delta<0$, $t\to\infty$ limit does not
obey this criteria. Of course, in this case, one may consider
$\Sigma_{n,\sigma}$ as an internal space. Still, however, one may
wonder if it is possible to  extend the solutions over $t=\infty$
region, and try to determine the maximal possible extensions and global
properties. It would also be interesting to understand the motion of
test particles on these time dependent backgrounds.

\section{Conclusions}

\hs{5}In this paper, we have constructed intersecting S-brane solutions of
$D=11$ supergravity theory. Our main result is that for S-branes there
is an analog of the harmonic function rule. It is remarkable that
using these rules, one can construct all standard intersecting
SM-brane solutions. It is also very peculiar that S-brane construction
rules cannot be applied to non-standard intersections and it seems
difficult to obtain explicit solutions corresponding to these
configurations.  By dimensional reduction and applying S
and T-duality transformations, one can also obtain all standard 
intersecting S-brane solutions of type IIA  and IIB string
theories. 

\
\

A crucial point which requires more work to understand is the physical
interpretations of the constants in the solutions. Especially, it would
be interesting to verify that the parameter $M_i$ is the energy of the
corresponding S-brane. Also, it is important to understand the
relation between $t_i$ and the moments that S-branes
exist. A careful definition of near brane and asymptotic regions
is very crucial in fixing these constants. It is 
interesting to recall that some intersections have
different asymptotic behavior depending on the constants
characterizing the  solution.

\
\

Unfortunately, as other examples of S-brane solutions,
intersecting backgrounds turn out to be singular. 
As discussed in \cite{s1} there are several possibilities
for these singularities. Perhaps the most fortunate fate 
is the one in which they are smoothed out by non-perturbative $\alpha
'$ or $g_{s}$ effects.

\
\

The fundamental
problem about these solutions is the question of their stability. 
Since they do not preserve any supersymmetry, one needs different
arguments to claim stability. 
The presence of a  singularity is a bad sign. On the other hand, the fact
that spacelike backgrounds corresponding to stable supersymmetric
configurations (and not the others) can easily be obtained by an
analog of the harmonic function rule gives some hope about the
stability of these solutions.  

\section*{Appendix}

\
\

The bosonic action of the 11-dimensional supergravity can be written as
\be
S=\int d^{11}x\sqrt{-g}(R-\frac{1}{2.4!}F^2)+\textrm{WZ}
\ee
where the WZ-term is proportional to $\int F\wedge F \wedge A$
and will not be important in this paper. The equations of motion are 
given by
\bea\label{eins}
R_{AB}&=&\frac{1}{2.3!}F_{ACDE}F_{B}{}^{CDE}-\frac{1}{6.4!}g_{AB}F^{2},\\
d*F&\sim&F\wedge F.
\eea
We are interested in configurations where $F\wedge F=0$ and thus the
four-form field equations reduce to $dF=d*F=0$. 

\
\

We consider a metric of the following form
\be
ds^2=-e^{2A}dt^2\,+\,\sum_k\,e^{2C_k}\,ds_k^2\,+\,e^{2D}\,d\Sigma_{n,\sigma}^2
\ee
where $ds_k^2$ (for $k=1,2..$) is the metric on the $d_k$ dimensional flat
space. $d\Sigma_{n,\sigma}^2$ is the metric of the
$n$-dimensional unit sphere $\sigma=1$, unit hyperbola $\sigma=-1$ or
flat space $\sigma=0$.  The metric functions are
assumed to depend only on $t$. With respect to the the
orthonormal frame $E^{0}=e^{A}dt$, $E^{\alpha_k}=e^{C_k}dy^{\alpha_k}$
and $E^{\theta}=e^{D}e^{\theta}$, where $e^{\theta}$ is an orthonormal
frame on $\Sigma_{n,\sigma}$, the Ricci tensor can be calculated as 
\bea
R_{00} & = & e^{-2A}\left[-\sum_{k} d_k(C_k^{''}+ C_k^{'2} -C_k^{'}A^{'})-n(D^{''}+ D^{'2} -D^{'}A^{'})\right]\nonumber\\
R_{\alpha_i\beta_i} & = &
e^{-2A}\left[C_{i}^{''}+d_{i}C^{'2}_{i}-C_{i}^{'}A^{'}\,+\,\sum_{k\not=i}d_{k}C_{k}^{'}C_{i}^{'}\,+\,nD^{'}C_{i}^{'}\right]\delta_{\alpha_i\beta_i}\,\,\,i=1,2...\label{r}\\
R_{\theta_1\theta_2} & = & e^{-2A}\left[ D^{''}\,+\,nD^{'2}-D^{'}A^{'}\,+\,\sum_{k}d_kC_{k}^{'}D^{'}\right]\delta_{\theta_1\theta_2}\,+\,\sigma(n-1)e^{-2D}\delta_{\theta_1\theta_2}\nonumber
\eea
where $'$ denotes differentiation with respect to $t$. One can
simplify Ricci tensor by fixing $t$-reparametrization invariance so
that
\be
A=\sum_{k}d_{k}C_{k}\,+\,nD.
\ee
With this gauge choice, the Ricci tensor becomes
\bea
R_{00} & = & e^{-2A}\left[-A^{''}+A^{'2}-\sum_{k}d_{k}C_{k}^{'2}-nD^{'2}\right]\nonumber\\
R_{\alpha_i\beta_i} & = & e^{-2A}
\left[C_{i}^{''}\right]\delta_{\alpha_i\beta_i}\,\,\,i=1,2...\\
R_{\theta_1\theta_2} & = & e^{-2A}\left[D^{''}\right]\delta_{\theta_1\theta_2}\,+\,\sigma(n-1)e^{-2D}\delta_{\theta_1\theta_2},\nonumber
\eea
and the curvature scalar is
\be
R=e^{-2A}\left[2A^{''}-A^{'2}+\sum_kd_kC_k^{'2}\,+\,n\,D^{'2}\right]\,+\sigma
n(n-1)\,e^{-2D}.
\ee
Right hand side of (\ref{eins}) suggests the introduction of functions
$f_i$ for each brane which is 
\be\label{f}
f_i\,=\,-\,A\,+\,\sum \,d_k\,C_k\,+\,nD,
\ee
where the summation is over transverse directions to that brane.
The $\sigma$ term in (\ref{r}) implies the definition of a function
$g$ as
\be\label{g}
(n-1)\,g\,=\,A-D.
\ee
After these definitions, it is possible to diagonalize the spatial
components of Einstein equations by parametrizing the metric functions
in terms of  $f_i$ and $g$ so that 
\bea\label{eqf}
f_i''&=&q_i^2\,e^{-2f_i},\\
g''&=&-\sigma(n-1)e^{2(n-1)g}.\label{eqg}
\eea
This parametrization implies (\ref{overall}), (\ref{m2h}) and
(\ref{m5h}) where $H_i=e^{2f_i}$ and $G_{n,\sigma}=e^{-2(n-1)g}$. It is
easy to see that (\ref{eqf}) and (\ref{eqg}) are integrable and 
give (\ref{H}) and (\ref{G}). The form of the integration constants
in (\ref{H}) and (\ref{G}) together with the relation (\ref{sum}) are
fixed using the $(00)$ component of the Einstein equations
(\ref{eins}).

\end{document}